\documentclass[prd,superscriptaddress,amsmath,twocolumn]{revtex4-2}
\usepackage{graphicx}
\usepackage{float}
\usepackage{tensor}
\usepackage{epstopdf,cancel}
\usepackage{epsf,latexsym,bbm,euscript}
\usepackage{amssymb,amsmath}
\usepackage{mathtools} 
\usepackage{times,graphics}
\usepackage{soul,xcolor}
\usepackage{mathtools}

\usepackage{enumitem}

\def\6{{\langle}}
\def\9{{\rangle}}
\newcommand{\defeq}{\vcentcolon=}

\newcommand{\be}{\begin{equation}}
\newcommand{\ee}{\end{equation}}
\newcommand{\ba}{\begin{eqnarray}}
\newcommand{\ea}{\end{eqnarray}}

\newcommand{\rA}{{\mathrm{A}}}
\newcommand{\mS}{{\mathrm{S}}}

\def\half{{\tfrac{1}{2}}}

\def\pad{{\partial}}

\def\sg{\textsl{g}}

 \def\eR{\EuScript{R}}
\def\cO{\mathcal{O}}

\def\mE{\mathfrak{E}}
\def\maT{\mathfrak{T}}

\def\iPhi{\mathit{\Phi}}

\usepackage{url,hyperref}
\hypersetup{colorlinks,linkcolor={blue!55!black},citecolor={red!45!black},urlcolor={blue!45!black},breaklinks=true}

\begin{document}

\title{Inaccessibility of   traversable  wormholes}

\author{Daniel R.\ Terno}
\affiliation{School of Mathematical and Physical Sciences, Macquarie University, Sydney, New South Wales 2109, Australia}

\begin{abstract}
Wormhole solutions to the equations of general relativity have some spectacular local and global properties. As these unusual features are not explicitly forbidden by known physics, wormholes are considered in various astrophysical and cosmological scenarios. The paradigmatic  traversable wormhole models are  described by static spherically-symmetric  Ellis--Morris--Thorne   and    Simpson--Visser metrics. We show that no dynamical solution of the semiclassical Einstein equations can have these metrics as their static limit. On the other hand, possible static limits of the dynamical solutions are not traversable. Moreover, they lead to violation of a quantum energy inequality that bounds violations  of the null energy condition by quantum fields. This conclusion does not depend on specific properties of fields that may be proposed for  wormhole construction. As a result, spherically-symmetric wormholes cannot exist in semiclassical gravity.
      \end{abstract}

\maketitle

\section{Introduction}
Wormhole solutions in general relativity \cite{V:96,L:17,K:18} are known for   almost as long the Schwarzschild metric. However, their status is different. The
black hole paradigm    explains all current observations of dark massive  ultra-compact objects. 
Wormholes are  just one of the alternative models. As such they aim to describe  astrophysical black holes without causing conceptual problems inherent in the notions of  an event horizon and singularity \cite{bh-map,CP:19,BS:21}.  Even the  revival of interest in   wormholes as traversable shortcuts between separated spacetime regions 
had sci-fi as its primary motivation \cite{MT:88,JTFT:15}.

Several   of  wormhole features, even if not forbidden by the laws of physics,  are unusual enough to  make  their existence  unlikely \cite{JTFT:15}.    The null energy condition (NEC) is the weakest   energy requirement  that is used in general relativity. The NEC is satisfied by normal classical matter \cite{HE:73,M_MV:17}. Quantum field theory permits its violations that are, however, constrained by quantum energy inequalities \cite{F:17,KS:20}.  NEC violation is a generic and universal feature of (traversable)
 wormholes \cite{HV:98}.  Creation of a wormhole implies a change in the topology of space, and using a pair of wormholes is a simple way to generate closed timelike loops, i.e. to create a time machine  \cite{V:96,L:17,K:18,JTFT:15}.

On the other hand, the NEC could be sufficiently violated in the early Universe to enable formation of wormholes \cite{R:93,V:96,L:17}. Its violation is a necessary condition for formation of a trapped spacetime region, i.e. a physical black hole (PBH) \cite{F:14}, in finite time of a distant observer \cite{HE:73,MMT:21}. Moreover, the necessary NEC violation can be effected by a non-phantom fermionic matter \cite{BKR:21,KZ:22}.   Topology changes are expected to occur in quantum gravity and are a basic component of the path integral approach to it \cite{L:17,H:09}. These arguments provide additional impetus for search for   astrophysical wormholes, using both electromagnetic and gravitational   radiation \cite{bh-map,CP:19,BS:21}.

 The original static wormhole solutions were characterized using the embedding diagrams and explicit description of the two spatial sheets that are connected at the wormhole's throat \cite{E:73,MT:88,MTY:88}.
  The invariant characterization of the throat that is valid for generic wormholes  identifies it as an outer marginal trapped surface subject to additional conditions \cite{HV:98,H:09}. This allows to describe dynamical wormholes \cite{H:09,SH:02}, and also to apply the self-consistent analysis of black hole horizons \cite{BMMT:19}.
    In case of spherical symmetry it allows an exhaustive description of the admissible solutions \cite{MMT:21}.

Using properties of these solutions we find that
 the standard
static traversable wormhole (TWH) solutions  are not   static limits of dynamical solutions. Moreover,
the admissible static limits are not only non-traversable wormholes, but violate the quantum energy inequalities (QEIs), making their introduction in semiclassical physics self-contradictory.

This article is organized as follows. In the next section we review the semiclassical physics of spherically-symmetric horizons. Sec.~\ref{TWH-s} reviews the basic properties of the static wormhole solutions, and positions the Ellis--Morris-Thorne  and the Simpson--Visser metrics within a general scheme of self-consistent solutions.  Sec.~\ref{dyn} contains the main original results of this work and presents the dynamic wormhole solutions and their static limits. The results, their implications and future directions are discussed in Sec.~\ref{disc}.

We use the $(-+++)$ signature and use the Planck units. Derivatives of a function of a single variable are marked with a prime, e.g.  $r'_\sg(t)\equiv dr_\sg/dt$. Derivative with respect to the proper time $\tau$ are denoted by the dot, $\dot R\equiv dR/d\tau$.

\section{The set-up}
Properties of TWHs are usually investigated by first designing a metric with the desired properties. This is possible if   the classical notions and semiclassical and/or modified Einstein equations are applicable. Then the Einstein  equations
  \begin{align}
	\tensor{G}{_\mu_\nu} = 8 \pi \tensor{T}{_\mu_\nu} \equiv 8 \pi \6 \tensor{\hat{T}}{_\mu_\nu} \9_\omega.
	\label{eq=1}
\end{align}
are reverse-engineered  to determine the energy-momentum tensor (EMT) that is the source of this  geometry \cite{L:17,SVS:21}.
Here the Einstein tensor $G_{\mu\nu}\defeq R_{\mu\nu}-\half \sg_{\mu\nu}\eR$, where $\tensor{R}{_\mu_\nu}$ and $\eR$ are the Ricci tensor and Ricci scalar, respectively,  is equated with the effective EMT. The latter is an expectation value  of the renormalized EMT operator plus all additional curvature terms. They appear, e.g., as a result of the renormalization procedure or are derived from the Lagrangian of  an effective field theory of gravity \cite{BD:82,B:04,DH:15}. We do not make any assumptions about nature of the fields or of the state $\omega$.

A general four-dimensional spherically symmetric metric in Schwarzschild coordinates is given by \cite{C:92,F:15}
\begin{align}
	ds^2 = -e^{2h(t,r)}f(t,r)dt^2+f(t,r)^{-1}dr^2+r^2d\Omega_2 , \label{eq:metric}
\end{align}
 These coordinates provide geometrically preferred foliations with respect to Kodama time, which is derived from a natural divergence-free vector field \cite{F:15,AV:10}. 
 Invariantly-defined Misner\textendash{}Sharp  mass \cite{F:15,MS:64} $C(t,r)/2$ allows to write
\begin{align}
	f(t,r) \defeq 1 - C/r \defeq \partial_\mu r \partial^\mu r. \label{eq:MS}
\end{align}
 In asymptotically flat spacetimes the coordinate $t$ is the physical time of a distant static observer (Bob). However, our results do not depend on this interpretation.

Using the advanced null coordinate $v$ the metric takes the form
\be
ds^2=-e^{2h_+(v,r)}f(v,r)dv^2+2e^{h_+(v,r)}dvdr +r^2d\Omega_2 \label{eq:metv},
\ee
where  $f=1-C_+(v,r)/r$ and invariance of the Misner--Sharp mass ensures $C_+(v,r)\equiv C\big(t(v,r),r\big)$. The coordinates are related via
\be
dt=e^{-h}(e^{h_+}dv-f^{-1}dr).
\ee

In all foliations that respect spherical symmetry \cite{FEFHM:17} components of the apparent horizon (a 3D boundary of the trapped region) \cite{HE:73,F:15,MMT:21} coincide  with the roots of
\be
f(t,r)=0. \label{hor0}
\ee
 The Schwarzschild radius $r_\sg(t)$, being the largest root, corresponds to the outer apparent horizon \cite{F:15,MMT:21}.
For the wormhole solutions $r_\sg(t)$ is the only root on each sheet of the radial coordinate. It corresponds to the throat of the wormhole, and the circumferential radius is restricted to the range $r_\sg\leqslant r<\infty$.
Hence constructing solutions of the Einstein equations that satisfy Eq.~\eqref{hor0} is the first step in developing wormhole solutions.

The self-consistent approach to black holes jointly identifies the forms of the EMT and of the metric functions $h$ and $C$ in the vicinity of the apparent horizon \cite{BMMT:19,MMT:21}. It is adopted here to generate the wormhole solutions.
Two requirements \cite{BMMT:19,MMT:21} allow to describe all potential geometries in the vicinity of $r_\sg$.
First,  Eq.~\eqref{hor0} is required to have a solution for  $t_\mS\leqslant t\leqslant t_*<\infty$ for some finite $t_\mS$. This allows for formation and a possible closure of the wormhole. It is important to note that despite the Misner\textendash{}Sharp mass invariance, having solutions of $f(v,r)=0$ or $f(v,u)=0$, where $v$ and $u$ are the advanced and the retarded null coordinates, respectively, does not imply that a TWH forms at finite time of Bob.

Second, the throat at $r=r_\sg$ is a regular 2-surface in a sense that the curvature scalars that are constructed from polynomials of components of the Riemann tensor are finite. Apart from a minimal compliance with the cosmic censorship conjecture it is also part of the requirements that ensure traversability of the wormhole \cite{L:17,MT:88}. We   use  two quantities that can be obtained directly from EMT components:
\begin{align}
	\tilde{\mathrm{T}}\defeq \tensor{T}{^\mu_\mu}, \qquad \tilde{\maT} = \tensor{T}{_\mu_\nu} \tensor{T}{^\mu^\nu} .
\end{align}
The Einstein equations relate them to the curvature scalars as $\tilde{\mathrm{T}} \equiv - {\eR}/8\pi$ and $\tilde{\mathfrak{T}} \equiv \tensor{R}{^\mu^\nu}\tensor{R}{_\mu_\nu}/64\pi^2$. For the spherically-symmetric solutions finite values of these scalars as $r\to r_\sg$ ensure that all independent scalar invariants  are finite \cite{MMT:21}.

It is convenient to introduce the effective EMT components
\begin{align}
	\tensor{\tau}{_t} \defeq e^{-2h} \tensor{T}{_t_t}, \qquad \tensor{\tau}{^r} \defeq \tensor{T}{^r^r}, \qquad \tensor{\tau}{_t^r} \defeq e^{-h} \tensor{T}{_t^r} . \label{eq:mtgEMTdecomp}
\end{align}
Then the  Einstein equations  for the components $\tensor{G}{_t_t}$, $\tensor{G}{_t^r}$, and $\tensor{G}{^r^r}$  are
\begin{align}
	\partial_r C &= 8 \pi r^2 \tensor{\tau}{_t} / f , \label{gtt} \\
	\partial_t C &= 8 \pi r^2 e^h \tensor{\tau}{_t^r} , \label{gtr} \\
	\partial_r h &= 4 \pi r \left( \tensor{\tau}{_t} + \tensor{\tau}{^r} \right) / f^2 . \label{grr}
\end{align}

To ensure the finite values of the curvature scalars it is sufficient to work with
\begin{align}
     \mathrm{T} \defeq (\tensor{\tau}{^r} - \tensor{\tau}{_t}) / f , \quad \mathfrak{T} \defeq \big( (\tensor{\tau}{^r})^2 + (\tensor{\tau}{_t})^2 - 2 (\tensor{\tau}{_t^r})^2 \big) / f^2 ,
     \label{eq:TwoScalars}
\end{align}
where the contribution of $T^\theta_{~\theta}\equiv T^\phi_{~\phi}$ is disregarded, and then  to verify that  the resulting metric functions do not introduce further divergences \cite{MMT:21}.
Thus   the three effective EMT components either diverge, converge to   finite limits or converge to zero   in such a way that  the above combinations are finite. One   option is the scaling
\begin{align}
	\tensor{\tau}{_t} \sim f^{k_E}, \qquad \tensor{\tau}{^r} \sim f^{k_P}, \qquad \tensor{\tau}{_t^r} \sim f^{k_\Phi} , \label{tauS}
\end{align}
for some powers $k_a>1$, $a=E,P,\Phi$. Another involves convergence or divergence with the same $k\leqslant 1$. For PBHs only solutions with $k=0,1$ are relevant.


 Solutions of the $k=0$ class satisfy
\begin{align}
	\tensor{\tau}{_t} \to \tensor{\tau}{^r} \to - \Upsilon^2(t), \qquad \tensor{\tau}{_t^r}\to\pm \Upsilon^2(t),
\end{align}
as $r \to r_\sg$. The negative sign of $\tau_t$ and thus of $\tau^r$ is necessary  to obtain the real-valued solutions of Eqs.~\eqref{gtt}--\eqref{grr}. The leading terms of the metric functions   near the outer apparent horizon are
\begin{align}
		C &= r_\sg - 4 \sqrt{\pi} r_\sg^{3/2} \Upsilon \sqrt{x} + \mathcal{O}(x) , \label{eq:k0C} \\
		h &= - \frac{1}{2}\ln{\frac{x}{\xi}} + \mathcal{O}(\sqrt{x}) , \label{eq:k0h}
\end{align}
where $\xi(t)$ is determined by the choice of time variable, and the higher-order terms are matched with the higher-order terms in the EMT expansion.  Eq.~\eqref{gtr} must then hold identically. Both sides contain terms that diverge as $1/\sqrt{x}$, and their matching results in the consistency condition
\begin{align}
	r'_\sg/\sqrt{\xi} =   4 \epsilon_\pm\sqrt{\pi r_\sg} \, \Upsilon , \label{eq:k0rp}
\end{align}
where $\epsilon_\pm = \pm 1$ corresponds to the expansion and contraction of the Schwarzschild sphere, respectively.

 Geometry near and across the Schwarzschild sphere is conveniently expressed \cite{BMMT:19,MMT:21} in $(v,r)$ coordinates for  $r'_\sg<0$, and in $(u,r)$ coordinates (where $u$ is the retarded null coordinate) for $r'_\sg>0$. The extended solutions describe an evaporating PBH ($r'_\sg<0$) and an expanding white hole ($r'_\sg>0$).
Vaidya metrics (with $M'(v)<0$ and $M'(u)>0$, respectively),  are examples of such objects belonging to the $k=0$ class.

Details of $k=1$ solutions are given in Sec.~\ref{dyn} and Appendix \ref{a:k=1}. Both $k=0$ and $k=1$ solutions satisfy
\be
\lim_{r\to r_\sg}e^h f=|r'_\sg|, \label{finT}
\ee
that ensures a finite infall time also according to a distant Bob \cite{MMT:21,T:20}.

The limiting form of the $(tr)$ block of a $k=1$ EMT as $r\to r_\sg$ is
\begin{align}
	\tensor{T}{^a_b} \approx \begin{pmatrix}
		\Upsilon^2/f & -\epsilon_\pm e^{-h}\Upsilon^2/f^2 \vspace{1mm}\\
		\epsilon_\pm e^h  \Upsilon^2 & -\Upsilon^2/f
	\end{pmatrix},   \label{emt0}
\end{align}
$a,b=t,r$. According to a static (outside) observer the local energy density, pressure and flux diverge as $r\to r_\sg$.

In addition to the usual list of requirements that make a  wormhole traversable,   experience with PBHs indicates another necessary feature: absence of strong firewalls, i.e. of divergent negative energy density and/or pressure and flux in the frame of a travelling observer (Alice) \cite{DMT:22,MMT:21}. These firewalls may occur even if the curvature scalars are finite. While  they indicate that the apparent horizon of a PBH is a surface of intermediate curvature singularity, a sufficiently strong firewall  leads to a divergent  integrated energy density, violating the QEIs. In particular, along a timelike geodesic $\gamma$  with a tangent four-vector $u^\mu_\rA$   the local energy density is
\begin{align}
	\rho_\mathrm{A}\defeq\6 \hat{T}^\mathrm{ren}_{\mu\nu} \9_\omega u_\rA^\mu u_\rA^\nu.
\end{align}
where the expectation of the renormalized EMT is evaluated on an arbitrary Hadamard state \cite{BD:82,KS:20} $\omega$.
The total integrated energy that is smeared by a sampling function $\wp(\tau)$  with a compact support. It can be taken to be $\wp\cong 1$ for an arbitrarily large fraction of the domain with $\wp>0$. Then
\begin{align}
	\int_\gamma d\tau \wp^2(\tau)\rho(\tau)\geqslant -B(\gamma, \eR,\wp), \label{qei-KO}
\end{align}
where $B>0$ is a bounded function that depends on the trajectory, the Ricci scalar, and the sampling function \cite{KO:15}. Violation of this bound by a particular solution  indicates its impossibility in semiclassical gravity. While integrated energy densities in case of the PBH firewalls are finite, we will see that some potential wormhole solutions lead to violation of this QEI.

We illustrate the issue by   considering $k=0$ solutions.  For an incoming Alice the energy density in her frame remain finite. It is $\rho_\rA\propto r'_\sg/\dot R$ at the apparent horizon \cite{MMT:21}.  On the other hand, for an outgoing observer outside a PBH
 \be
 \rho_\rA=\frac{\dot R^2}{4\pi r_\sg X}+\cO(1/\sqrt{X}),
 \ee
where $X\defeq R(\tau)-r_\sg\big(T(\tau0\big)$. The energy density is obtained by using the expression for the EMT of $k=0$ solutions (Eq.~\eqref{emt0}) and the normalization of the four-velocity in the form
\be
\dot T=\frac{\sqrt{\dot R^2+F}}{e^H F}\approx \frac{|\dot R|}{|r'_\sg|},
\ee
where $F=f(T,R)$ and $H=h(T,R)$. For a PBH this divergence is largely a curios observation (observers that may exit the so-called quantum ergosphere cannot have $\dot R>0$ at the apparent horizon \cite{MMT:21,DMT:22}). However,
if this metric describes the neighbourhood of a wormhole throat for $r \gtrsim r_\sg$, then an exiting Alice will have $\dot R>0$ at the throat, and such a divergence contradicts traversability.

\section{Standard static TWHs} \label{TWH-s}

In horizonless spacetimes it is often convenient to represent a spherically-symmetric metric as
\be
ds^2 = -e^{2\iPhi}dt^2+f(t,r)^{-1}dr^2+r^2d\Omega_2 , \label{eq:mwh}
\ee
 where $\iPhi\equiv h+\half\ln f$. Requiring $\sg_{tt}(t,r_\sg)$ to be finite while $f\to 0$ implies that in addition to  Eq.~\eqref{grr} the equation
\be
\pad_r h\approx -\frac{\pad_r f}{2f}\approx-\frac{\lambda}{2x}, \label{drh}
\ee
holds as well, where we kept only the divergent terms. The final expression above is a leading term in the expansion in powers of $x\defeq r-r_\sg$ of $f\approx \phi(t)x^\lambda$, where $\phi(t)$ is some  function.

 After
 substituting Eq.~\eqref{drh} in Eq.~\eqref{grr} and using Eq.~\eqref{gtt} we obtain a local algebraic relation between the metric functions mass and the  EMT component. If $\iPhi=0$ then
the relation $ \tau^r=-Cf/(8\pi r^3)$ holds exactly. On the other hand,  the expansion
\be
\tau^r=-(8\pi r_\sg^2)^{-1}f+\ldots \label{eq4}
\ee
 near the Schwarzschild radius is valid for a general finite $\iPhi$.

For static wormholes the physical distance allows to introduce the coordinate $l$,  $-\infty<l(r)<\infty$, that describes both sides of the bridge via $dl=\pm dr/\sqrt{f}$.
The regular metric functions  have series expansions,
\be
C=b_0+b_1 x +b_2x^2+\ldots, \qquad \iPhi=\phi_1 x+\phi_2 x^2+\ldots,
\ee
where $x\defeq r-b_0$, and we absorbed the constant term in $\iPhi$ by redefining the time. The leading coefficient $b_1$ must satisfy  $b_1\leqslant 1$ as to make the throat at $r_\sg\equiv b_0$ a marginally trapped surface. Then
\be
f(r)=\frac{(1-b_1)x}{b_0}+\cO(x^2).
\ee
and
\be
h=-\half \ln\frac{x}{\xi}+\left(\phi_1+\frac{1-b_1+b_0b_2}{2b_0(1-b_1)}\right)x+\ldots,
\ee
where
\be
\xi\equiv \frac{b_0}{1-b_1}.
\ee
This form allows for a more convenient comparison with PBH solutions. At the throat
\be
\rho(r_\sg)=\frac{b_1}{8\pi b_0^2}, \qquad p(r_\sg)=-\frac{1}{8\pi b_0^2}.
\ee
Expansion of $\tau_t$, $\tau^r$ and $\tau_t^{~r}$ in terms of $x$ shows that for $b_1\neq 0$ these solutions belong to $k=1$ class.

The standard Ellis--Morris--Thorne metric \cite{E:73,MT:88} corresponds to $\iPhi=0$ and
\begin{align}
C&=b_0^2/r, \\
 h&= -\half\ln 2x/b_0+\cO(x),
 \end{align}
 with the throat at $r_\sg=b_0$, and thus
\be
b_1=-1, \qquad b_2= 1/b_0.
\ee


 The Simpson--Visser metric \cite{SV:19} interpolates between the Schwarzschild black hole and TWHs,
\be
ds^2=-\left(1-\frac{2m}{\sqrt{\eta^2+a^2}}\right)dt^2+\frac{d\eta^2}{1 -\frac{2m}{\sqrt{\eta^2+a^2}} } +(\eta^2+a^2)d\Omega_2, \label{svm}
\ee
where $a$ is a parameter. The Misner\textendash{}Sharp mass is
$C=2m+a^2(r-2m)/r$, and for $a\geqslant 2m$ the throat is located at $r_\sg=a$.

Expanding the metric functions near the throat we find for $a>2m$
\begin{align}
C&=a+\frac{4m-a}{a}x+\cO(x^2), \\
h&=-\frac{1}{2}\ln\frac{2(a-2m)}{a}x+\cO(x),
\end{align}
and for $a=2m$ (the one-way wormhole \cite{SV:19})
\begin{align}
C&=r+\frac{2}{a}x^2+\cO(x^3), \\
h&=- \ln\frac{\sqrt{2}x}{a}+\cO(x).
\end{align}


Both Ellis--Morris--Thorne and Simpson--Visser metrics belong to the $k=1$ class.

\section{Dynamical wormhole solutions and their limits} \label{dyn}

  There are no static $k=0$ solutions, as in this case the scalar $\tilde\maT$ cannot be finite. On the other  hand,   static solutions are not only possible in the class of  $k=1$,
\begin{align}
	\tensor{\tau}{_t} \to E(t) f, \qquad \tensor{\tau}{^r} \to P(t) f, \qquad \tensor{\tau}{_t^r} \to \Phi(t) f,  \label{eq:k1taus}
\end{align}
but the two most popular static TWH metrics belong to it.

The energy density $\rho(t,r_\sg)=E$ and the pressure $p(t,r_\sg)=P$ are finite at the Schwarzschild radius. (This is also true in the proper reference frame of a static observer).
For $E<(8\pi r_\sg^2)^{-1}$   the resulting metric functions are
 \be
C=r_\sg(t)+8\pi E r^2_\sg x+\ldots,   \label{regas}
\ee
and
\be
h=-\ln \frac{x}{\xi(t)}+\cO(\sqrt{x})+\ldots,    \label{hgas}
\ee
for some $\xi(t)>0$. Consistency of the Einstein equations results in
\be
P=E-\frac{1}{4\pi r_\sg^2}, \qquad \Phi=\pm\left(\frac{1}{8\pi r_\sg^2}-E\right). \label{k=0rel}
\ee
Eq.~\eqref{gtr} implies now
\be
\qquad r'_\sg=8\pi\Phi\xi r_\sg.
\ee

A static $k=1$ configuration can be reached from the solution where the energy density at $r_\sg$ takes its maximal possible value $E=1/(8\pi r_\sg^2)$ \cite{MT:21}. While only such extreme $k=1$ solutions can describe PBHs \cite{MT:21}, there is no such restriction on potential wormhole solutions. In the extreme limit $P=-E$, $\Phi=0$, and
\begin{align}
	C(t,r)=r+ c_{32}(t)x^{3/2}+c_2(t) x^2 +\cO(x^{3/2}), \label{fk1}
\end{align}
for some coefficient $c_{32}(t)<0$,
and
\begin{align}
	h=-\tfrac{3}{2}\ln (x/\xi)+\mathcal{O}(\sqrt{x}), \label{hk1}
\end{align}
while Eq.~\eqref{gtr} implies
\be
	r'_\sg=\pm |c_{32}|\xi^{3/2}/r_\sg. \label{rder1}
\ee
The static limit is possible if as $t\to t_0$ the parameters $c_{32}(t)\to 0$ and $\xi(t)\to \xi_0$ (see Appendix~\ref{a:k=1} for details). In this case the static
 solution has $f=|c_2|x^2/r_\sg +\cO(x^{3/2})$ for some constant $c_2<0$, while $h$ given by Eq.~\eqref{hk1} with $\xi=\xi_0$.

As indicated by a different behavior of the function $h$ (a pre-factor $\frac{3}{2}$ instead of $\half$) the resulting static metric is different from the  standard TWH metrics.  This can be attributed to existence of  an additional constraint:  a  real solution with $f=0$ and finite $\tilde{\mathrm{T}}$ and $\tilde{\maT}$ also has a finite non-zero $g_{tt}$. This extra requirement cannot be satisfied dynamically while conforming to the two basic PBH conditions.

Indeed, if $0<|g_{tt}|<\infty$ then  Eq.~\eqref{eq4} implies, since $\mathrm{T}$ and $\maT$ are finite, that the solution  either belongs to the class $k=1$ or to one of the classes with $k_E>1$.  In the $k=1$ case we find $E=-P=1/(8\pi r_\sg^2)$, $\Phi=0$. Following through we then arrive to Eqs.~\eqref{fk1} and \eqref{hk1},  contradicting Eq.~\eqref{drh} and thus the initial assumption. Looking at the observable quantities, we find, e. g. ,
that for the Ellis--Morris--Thorne metric $E=+P=-1/(8\pi r_\sg^2)$, which is impossible in    dynamical solutions.

Solutions with $k_E>1$ are possible. Consistent dynamical solutions exist for half-integer values of $k_E\geqslant 2$ and $k_P=k_{\Phi}=1$ \cite{T:20}. However, in addition to presence of a strong firewall for some of the geodesic observers, they do not have a static limit: the Ricci scalar $\eR(t, r_\sg)$   diverges when $r'_\sg=0$ (see Appendix~\ref{a:k>1} for details).

So far we have seen that the standard TWH solutions are not the static limits of some allowed semiclassical  solutions of the Einstein equations. We now demonstrate that  the static limits of the allowed $k=1$  wormhole solutions are not traversable.

 For definiteness, consider an ingoing radial trajectory of Alice, $u_\rA=(\dot T,\dot R,0,0)$, $u_\rA^2=-1$. Energy conservation on a static background determines the radial velocity via $\dot R=-\sqrt{\mE^2-f(R)}$, where $\mE\geqslant 1$ is Alice's energy per unit mass at infinity. Using the EMT that is reverse engineered from the metric of Eqs.~\eqref{fk1} and \eqref{hk1} with $c_{32}=0$ and  $\xi=\xi_0$,  we find
 \be
 \rho_\rA=-\frac{3\mE^2}{8\pi r_\sg X}+\cO(\sqrt{X}),
 \ee
where $X(\tau)\defeq R(\tau)-r_\sg$ and  $\dot X =\dot R$. We   choose the sampling function $\wp=1$ in some vicinity of the throat and let $\wp\to 0$  still within the NEC-violating domain. As the trajectory
 passes through  $X_0+r_\sg\to r_\sg$ the lhs of Eq.~\eqref{qei-KO} behaves as
 \be
 \int_\gamma \wp^2\rho_\rA d\tau= \frac{3\mE}{8\pi r_\sg}\int_\gamma  \frac{1+\cO(\sqrt{X})}{X}dX\propto \log X_0\to-\infty,
 \ee
where we used $\dot R\approx -\mE$ in the vicinity of the Schwarzschild radius $r_\sg$. The right hand side of Eq.~\eqref{qei-KO} is finite, and thus the QEI is violated.

Similar firewalls occur also in dynamical solutions. For example, exiting through the contracting ($r'_\sg<0$) throat of $k=0$ wormhole with finite radial velocity $\dot R>0$ leads to $\rho_\rA\propto 1/X$, which is stronger than weak  firewalls on   trajectories that  cross apparent horizon of a PBH \cite{DMT:22}.



\section{Discussion} \label{disc}

The requirement of finite time of   formation according to a distant observer not only probes the constraints that ``the laws of physics place on the activities
of an arbitrarily advanced civilization" \cite{MTY:88}, but investigates the local implications of the topology change. The minimal regularity requirement not only enforces compliance with the cosmological censorship, but is a part of the traversability requirement.

 In spherical symmetry these two necessary assumptions are enough to produce an exhaustive description of potential  geometries. However, none of them leads in the standard static TWHs.

Impossibility of wormhole  formation in finite time is different from the asymptotic nature of, say, the Schwarzschild solution. Even if the apparent horizon may never form according to Bob \cite{MMT:21}, the classical black hole geometry provides an excellent description of the exterior, while approach to it is exponentially fast.
On the other hand, the defining feature of a wormhole is its being  a shortcut between   spacetime regions \cite{V:96,K:18}. Without a throat where the connection happens there is no wormhole. In the scenarios where a stationary limit can be reached the resulting wormhole is not traversable: Alice experiences  an infinite negative energy density. Moreover, similarly to some dynamical configurations this divergent negative energy violates  the QEI that prescribes a finite value to the integrated NEC violation.

   Various TWH solutions were designed under assumption that the semiclassical gravity is valid. Studies of the amount of the NEC violation, necessary matter content, etc, are performed within this framework. The self-consistent analysis of the semiclassical Einstein equations does not use any information about the matter content. Its main conclusion is that none of the possible spherically-symmetric wormhole solutions is a TWH.
   As a result, we have to accept  that existence of macroscopic spherically-symmetric wormholes requires not only a large-scale violation of the NEC and violation of the topology and chronology projection conjectures, but breakdown of semiclassical gravity.  It is remarkable that the analysis is based on local considerations, despite the global implications of wormhole existence.

   This work is limited to spherically-symmetric solutions. To answer the question whether existence of traversable wormholes  within the confines of known physics can be ruled out on the basis of local considerations  more general configurations, and in particular axially-symmetric scenarios should be considered.

\acknowledgments
 Useful discussions with Eleni Kontou, Robert Mann and Sebastian Murk are gratefully acknowledged. This work is supported by   the ARC Discovery project grant DP210101279.

\appendix

\section{Some properties of $k=1$ solutions}\label{a:k=1}
For the extreme solutions $E=1/(8\pi r_\sg^2)$.   Eq.~\eqref{k=0rel} then implies
 $\Phi=0$ and  $P=-E=- 1/(8\pi r_\sg^2)$. As a result $C(t,r)$ is given by Eq.~\eqref{regas}, and using the next order EMT expansion   leads to Eqs.~~\eqref{fk1} and ~\eqref{hk1}.
 A potentially divergent term in expansion of the Ricci scalar,
 \be
 \eR_\mathrm{div}=\frac{3}{2\sqrt{x}}\left(\frac{c_{32}}{r_\sg}-\frac{r_\sg r_\sg'}{c_{32}\xi^3}\right),
 \ee
 is identically zero due to Eq.~(L26). However, metrics with   $C-r$ starting with a higher power of $x$, such as $f=|c_2|x^2/r_\sg +\cO(x^{3/2})$ can be only static: the Ricci scalar is finite at the apparent horizon only if $r_\sg'=0$.

In the static case as $r\to r_\sg$ is
\begin{align}
	T_{ab} \approx \frac{1}{8\pi r_\sg x}\begin{pmatrix}
		- c_2\xi_0^3/r_\sg^2  & 0 \\
		0 &  1/(c_2 x)
	\end{pmatrix},
	\end{align}
while the leading order expansion of the 4-velocity of a free-falling Alice is
\be
u^\mu_\rA\approx \mE\left(\frac{r_\sg}{c_2 \xi_0^{3/2}\sqrt{X}},\pm 1\right),
\ee
where $\mE=\mathrm{const}$ is Alice's energy per unit mass at infinity.

\section{Some properties of $k_E>1$ solutions}\label{a:k>1}

Here we adopt the analysis of \cite{T:20}.
For the effective EMT components   with $k_a\geqslant 1$ the leading terms in the Einstein equations  become
\begin{align}
&\pad_r C \approx 8\pi r_\sg^2 {E(t)} {f^{k_E-1}}, \label{gttA1}\\
& \pad_t C \approx 8\pi r_\sg^2 e^h \Phi(t) f^{k_\Phi}, \label{gtrA1}\\
& \pad_r h\approx 4\pi r_\sg\big( E(t) f^{k_E-2}+P(t) f^{k_P-2}\big), \label{grrA1}
\end{align}
for some functions $E(t)$, $P(t)$ and $\Phi(t)$  and the powers $k_E,k_\Phi,k_P\geqslant1$.

 The leading terms of the Misner-Sharp mass are then
  \be
  C=r_\sg(t)+8\pi E r_\sg^2 x^{k_E}+\ldots.
  \ee
Solutions with $k_E=1$ and $k_P>1$ and/or $k_\Phi>1$ lead to a divergent Ricci scalar. However, solutions with $k_E\geqslant\tfrac{3}{2}$ are consistent. In this case
   \be
   f= {x}/{r_\sg}+\ldots.
   \ee

 Solutions with variable $r_\sg(t)$ impose via Eq.~(L8)  the logarithmic divergence of the function $h$, as it is necessary that $e^h\propto x^{-k_\Phi}$.  It can be realized only if $k_P=1$. Then
\be
  h=4\pi P r_\sg^2 \ln \frac{x}{\xi}, \qquad   4\pi P r_\sg^2=-k_\Phi.
   \ee
Expressing this solution in $(u,r)$ or $(v,r)$ coordinates leads to $k_\Phi=1$ and $\Phi=\pm 1/(8\pi r_\sg^2)$. These solutions are rather
  peculiar:   energy density vanishes at $r_\sg$ and the pressure and the flux are determined by the Schwarzschild radius.  A potentially divergent term in expansion of the Ricci scalar,
 \be
 \eR_\mathrm{div}=- \frac{1}{x}\left( \frac{1}{r_\sg}+\frac{r_\sg r_\sg'^2}{\xi^2}\right),
 \ee
and its leading coefficient is non-zero if $r_\sg'=0$.

\end{document}